\documentclass[twoside]{LCWS11}
\usepackage[latin1]{inputenc}
\usepackage[dvips]{graphicx,epsfig,color}
\usepackage{wrapfig,rotating}
\usepackage{amssymb,amsmath,array}
\usepackage{cite}
\begin{document}
\title{
Higgs Branching Fraction Study in ILC} 
\author{Hiroaki Ono$^1$ and Akiya Miyamoto$^2$
\vspace{.3cm}\\
1- Nippon Dental University School of Life Dentistry at Niigata \\
1-8 Hamaura-cho chuo-ku Niigata, Niigata - Japan
\vspace{.1cm}\\
2- High Energy Physics Research Organization \\
1-1 Oho Tsukuba, Ibaraki - Japan\\
}

\maketitle

\begin{abstract}
 Precise measurement of the Higgs boson properties are important issues for the International Linear Collider (ILC) project
 to understand the particles mass generation mechanism which strongly related to the coupling with the Higgs boson.
 Large Hadron Collider (LHC)~\cite{LHC} experiments exclude the large area of the predicted Higgs mass region 
 and their results indicate that Higgs boson mass will be light.
 Even if LHC discovers the Higgs like particle by the end of 2012,
 Higgs will be identified by the high precision measurement of the Higgs boson properties in ILC
 and also Higgs measurement verifies the correctness of standard model (SM) or gives some hints toward its beyond.
 In this study, we evaluate the measurement accuracies of Higgs branching fraction
 to the $H\to b\bar{b}$, $c\bar{c}$ and $gg$ at the center-of-mass energy of 250 and 350 GeV.
\end{abstract}

\begin{wraptable}{r}{0.48\textwidth}
 \begin{center}
  \caption{Expected SM Higgs BR at the Higgs mass of $120~{\rm GeV}$ in \texttt{PYTHIA}~\cite{PYTHIA}.}
 \label{table:BR}
  \begin{tabular}{|l|c|}
   \hline
   $H$ decay mode & $BR$s at $M_{H}=120~{\rm GeV}$\\
   \hline\hline
   $H \to b\bar{b}$         & 65.7\%\\
   $H \to WW^{*}$           & 15.0\%\\
   $H \to \tau^{+}\tau^{-}$ &  7.9\%\\
   $H \to gg$               &  5.5\%\\
   $H \to c\bar{c}$         &  3.6\%\\
   \hline
  \end{tabular}
 \end{center}
\end{wraptable}
 \section{Introduction}
\begin{figure}[htbp]
 \begin{center}
  \includegraphics[width=0.9\columnwidth]{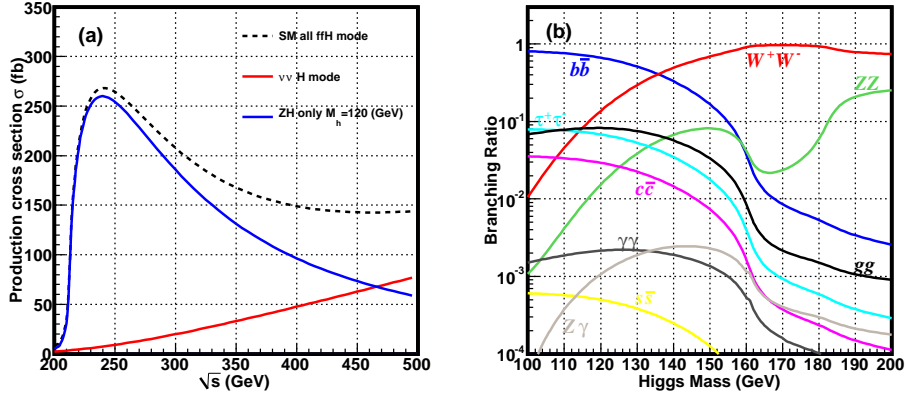}
  \caption{(a) Higgs production cross section as a function of CM energy and (b) SM predicted Higgs BR as a function of Higgs mass.}
  \label{fig:xsec_BR}
 \end{center}
\end{figure}

 Higgs branching ratio (BR) measurement is one of the important issue of the International Linear Collider (ILC) project,
 which is strongly related to the coupling strength with particles and reveal their mass generation mechanism.
 Even the Large Hadron Collider (LHC)~\cite{LHC} will discover the Higgs boson in a few year,
 ILC can confirm whether that is the standard model (SM) predicted one or not, and find some hints toward its beyond.
 LHC experiment accumulate the total integrated luminosity up to $5~\rm fb^{-1}$ by the end of 2011,
 LHC gradually exclude the heavy Higgs mass region and indicates the light Higgs
 from the combined results of ATLAS~\cite{ATLAS} and CMS~\cite{CMS} ($115 \leq M_{H} \leq 135~{\rm GeV}$).
 In this region, we obtain the maximum production cross section around the center-of-mass energy ($\sqrt{s}$) of 250 GeV
 and Higgs boson BR significantly varies depending on the Higgs mass,
 especially main decay channel shifts from $H\to b\bar{b}$ to $WW^{*}$ around the $M_{H}=140~{\rm GeV}$,
 as shown in Fig.~\ref{fig:xsec_BR} (a) and (b).
 In this study, we evaluate the measurement accuracies of the Higgs BRs
 of $H\to b\bar{b}$, $c\bar{c}$ and $gg$ with assuming the Higgs mass of 120 GeV and integrated luminosity (${\cal L}$) of $250~{\rm fb^{-1}}$, using the International Large Detector (ILD)~\cite{ILD} full simulation at $\sqrt{s}=250~{\rm GeV}$.
 In addition, we also consider the operation at the CM energy of 350 GeV,
 which increase the contribution of $W/Z$ fusion process but $Z/H$ will be boosted,
 for taking into account the energy staging option in ILC project.
 Considering the LHC results, we also evaluate the Higgs BR accuracy at the indicated mass region
 by extrapolating the $M_{H}=120~{\rm GeV}$ result.

 \section{Signal and Backgrounds}

 \begin{figure}[htbp]
  \begin{center}
   \includegraphics[width=0.9\columnwidth]{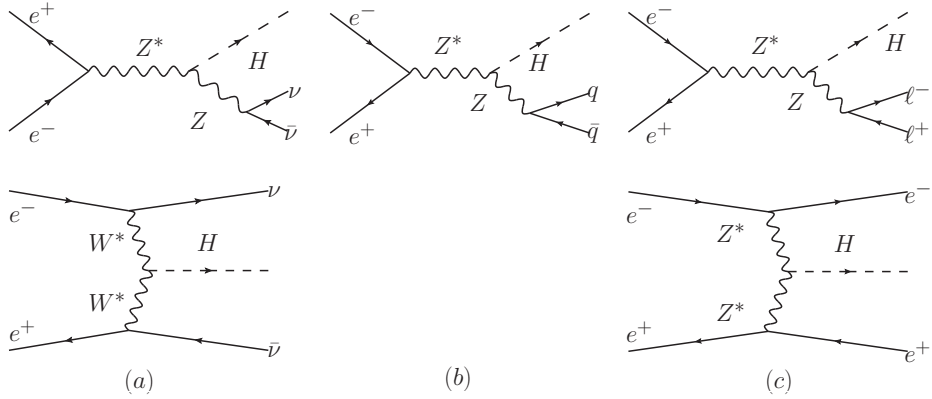}
   \caption{Higgs production processes categorized with $Z$ decay channels: (a) neutrino ($\nu\bar{\nu}H$), (b) hadronic ($q\bar{q}H$) and (c) leptonic ($\ell^{+}\ell^{-}H$).}
   \label{fig:hdiagram}
  \end{center}
 \end{figure}
 
 Figure~\ref{fig:hdiagram} shows the Higgs production diagrams and
 $ZH$ analysis procedures are categorized with the $Z$ decay channels;
 $ZH \to \nu\bar{\nu}H$ (Neutrino), $q\bar{q}H$ (Hadronic) and $\ell^{+}\ell^{-}H$ (Leptonic)
 which mainly form di-jet, four-jet and di-lepton+di-jet final states, respectively.
 We also consider the $W/Z$-fusion process for the neutrino and leptonic channels
 which has larger contribution at $\sqrt{s}=350~{\rm GeV}$.
 Assuming the $M_{H}=120~{\rm GeV}$, we obtain the maximum production cross section around the $\sqrt{s}=250~{\rm GeV}$
 through the Higgs-strahlung ($ZH$) process and mainly decays to $b$ quarks ($H\to b\bar{b}$).
 In order to maximize the Higgs production cross section,
 we employ the left-handed electron beam polarization; $(e^{+}, e^{-})=(+30\%, -80\%)$,
 at the CM energies of both 250 and 350 GeV with the integrated luminosity of $250~{\rm fb^{-1}}$.
 As background, we consider the following $2f$ and $4f$ final state SM Backgrounds:
 $e^{+}e^{-}\to W^{+}W^{-}$, $ZZ$ and $q\bar{q}$.
 In addition, we also take into account for the $e^{+}e^{-}\to t\bar{t}$ background only for the $\sqrt{s}=350~{\rm GeV}$.
 
 \section{Analysis Framework}

 Since $ZH$ final state forms multi-jet,
 thus jet clustering, jet energy resolution and quark flavor-tagging are crucial
 for the Higgs hadronic decay channels ($H \to b\bar{b}$, $c\bar{c}$ and $gg$) analysis.
 In order to achieve the best jet energy resolution,
 ILD adopt the Particle Flow Algorithm (PFA)~\cite{PFA},
 which can achieve the best jet energy resolution;
 charged tracks energy is measured by tracker instead of the calorimeter
 and only neutral particles energy is measured by calorimeter with avoiding the cluster overlapping and double counting.
 ILD detector design is well-suitable for the best PFA performance with adopting the finely segmented calorimeter,
 large tracker radius and strong magnetic field.
 For the simulation study, we use the \texttt{ilcsoft v01\_06}~\cite{ilcsoft} ILC common software package.
 At fist we generate the MC event samples with the \texttt{Whizard}~\cite{WHIZARD}.
 Then we perform the ILD full detector simulation with the \texttt{Mokka}~\cite{Mokka} package
 assuming the ILD detector model (\texttt{ILD\_00}).
 Generated hits are digitized and reconstructed with \texttt{Marlin}~\cite{Marlin} package
 and perform the PFA (\texttt{PandraPFA}~\cite{PFA}).
 We also employ the \texttt{LCFIVTX}~\cite{LCFI} flavor tagging package embedded in \texttt{Marlin}
 to identify the quark flavor of $H\to b\bar{b}$ and $c\bar{c}$.
 For the mass production of the full simulation samples,
 we use the GRID~\cite{GRID} resources for the ILD LOI study~\cite{ILD}
 and finally saved as ILC common file format (LCIO~\cite{LCIO}).

 \section{Event Reconstruction and Background Reduction}
 
  \subsection{Neutrino Channel ($\nu\bar{\nu}H$)}
   \subsubsection{Jet Reconstruction} 
   For the $\nu\bar{\nu}H$ channel analysis,
   we apply the jet reconstruction forcibly merged into the two-jet which comes from $H\to b\bar{b},~c\bar{c}$ and $gg$.
   
   \subsubsection{Background Reduction}

 In this channel, $ZZ \to \nu\nu qq$ and $WW \to \nu\ell qq$ will be the main backgrounds.
 To suppress these backgrounds,
 at first, we apply the missing mass cut; $80 < M_{miss} < 140~{\rm GeV}$
 to suppress the $ZZ$ to leptonic or hadronic decay backgrounds,
 since $M_{miss}$ should be consistent with the $Z$ mass in $Z\to \nu\bar{\nu}$ channel.
 Then we use the following kinematic variables cut: transverse momentum; $20 < P_{t} < 70~{\rm GeV}$, longitudinal momentum; $|P_{\ell}|<60~{\rm GeV}$ and maximum momentum; $P_{max}< 30~{\rm GeV}$ to suppress the $q\bar{q}$ background.
 Number of charged tracks cut; $N_{chd}>10$ well reduce the backgrounds including energetic leptons.
 We also apply the jet clustering $y$-value cuts; $0.2<Y_{12}<0.8$ and $Y_{23}<0.02$,
 which are the $y$-value thresholds from two- to one-jet or two- to three-jets
 and different of number of jets backgrounds are reduced with these cut.
 Finally we apply the di-jet mass cut; $80<M_{jj}< 130~{\rm GeV}$ which correspond to Higgs mass.
 After applying all the cuts, we apply the likelihood variable cut using following input variables;
 $M_{miss}$, number of reconstructed particles ($N_{PFO}$), $P_{max}$, $P_{l}$ and $M_{jj}$,
 and we select the $LR>0.375$ where we obtain the maximum signal significance.
 In order to optimize the cut positions for the analysis at $\sqrt{s}=350~{\rm GeV}$,
 we change the following variables cut positions:
 $50<M_{miss}<240~{\rm GeV}$, $10<P_{t}<140~{\rm GeV}$, $|P_{\ell}|<130~{\rm GeV}$ and $LR>0.15$.
 \begin{table}[htbp]
  \begin{center}
   \caption{Summary of the background reduction in $\nu\bar{\nu}H$ channel at the $\sqrt{s}=250$ and $350~{\rm GeV}$.}
   \begin{tabular}{l|r|r|r|r}
    \hline
     $\sqrt{s}$ (GeV)&\multicolumn{2}{c|}{250}&\multicolumn{2}{c}{350}\\
    \hline
    Cut  & Sig. & Bkg. & Sig. & Bkg.\\
    \hline\hline   
    Gen. & 19360 & 44827100 & 26307 & 20855900\\
    \hline
    All cuts & 6731 & 19058 & 12338 & 71918 \\
    \hline
    $LR$ cut & 4753 & 3593 &  9302 & 10029 \\
    \hline
    Significance (Eff.) & \multicolumn{2}{c|}{52.0 (24.5\%)}& \multicolumn{2}{c}{66.9 (35.4\%)}\\
    \hline
   \end{tabular}
  \end{center}
 \end{table}
 
 \subsection{Hadronic Channel ($q\bar{q}H$)}
   \subsubsection{Jet Reconstruction and Pairing}
   For $q\bar{q}H$ channel, we apply the four-jet reconstruction forcibly.
   Then we calculate following $\chi^{2}$ value to determine the $Z$ or $H$ jet pair candidates:
   \begin{equation}
    \chi^2 = \left( \frac{M_{12}-M_{H}}{\sigma_{H}} \right)^{2} + \left( \frac{M_{34}-M_{Z}}{\sigma_{Z}}\right),
   \end{equation}
   where $M_{12}$ and $M_{34}$ are reconstructed di-jet invariant masses and $\sigma_{Z/H}$ are the width of $Z$ and $H$ mass distribution.
   We select the minimum $\chi^{2}$ jet pairs as the best candidate of $Z$ and $H$.
   
   \subsubsection{Background Reduction}
   
   After the jet clustering and pairing, we apply the background reduction.
   At first we apply the $\chi^{2}$ cut to reduce the wrong combination pairs; $\chi^{2}<10$.
   Then we require the following cuts to suppress the leptonic events:
   the number of charged tracks cut; $N_{chd}>4$ and $y$-value cut $Y_{34}$ cut,
   which is a $y$-value threshold from three- to four-jet; $-\log{Y_{34}}<2.7$.
   As event shape cuts, we employ the following variables cut:
   thrust and its cosine of thrust angle; $thrust<0.9$, $|\cos{\theta_{thrust}}|<0.9$
   and the angle between the Higgs candidate jets;
   $105<\theta_{H}<160^{\circ}$ to suppress the $q\bar{q}$ backgrounds.
   Finally we apply the di-jet mass cut $M_{12}$ and $M_{34}$ and
   likelihood variable cut which is calculated with following input variables;
   thrust, $M_{Z}$, $M_{H}$, $\theta_{H}$.
   We select the likelihood variable cut as $LR>0.2$.
   For the $\sqrt{s}=350~{\rm GeV}$,  we optimize the cut position as follows;
   $thrust < 0.85$, $70<\theta_{H}<120^{\circ}$, $80 < M_{Z} < 100~{\rm GeV}$, $105 < M_{H} < 130~{\rm GeV}$
   and $LR>0.1$.
   
   \begin{table}[htbp]
    \begin{center}
     \caption{Summary of the background reduction in $q\bar{q}H$ channel at the $\sqrt{s}=250$ and $350~{\rm GeV}$.}
     \begin{tabular}{l|r|r|r|r}
      \hline
      $\sqrt{s}$ (GeV)&\multicolumn{2}{c|}{250}&\multicolumn{2}{c}{350}\\
      \hline
      Cut  & Sig. & Bkg. & Sig. & Bkg.\\
      \hline\hline   
      Gen. & 52507 & 44827100 & 36099 & 21222700\\
      \hline
      All cuts & 16350 & 411785 & 9447 & 44400 \\
      \hline
      $LR$ cut & 13726 & 166807 &  8686 & 25393 \\
      \hline
      Significance (Eff.) & \multicolumn{2}{c|}{32.3 (26.1\%)}& \multicolumn{2}{c}{47.1 (24.1\%)}\\
      \hline
     \end{tabular}
    \end{center}
   \end{table}
   
   \subsection{Leptonic Channel ($\ell^{+}\ell^{-} H$)}
  
  For the $\ell^{+}\ell^{-} H$ channel analysis, at first we identify the di-lepton,
  then we apply the di-jet reconstruction forcibly for remaining particles.
  
   \subsubsection{Di-lepton Identification}
   
   We apply the following di-lepton identification for electrons and muons
   from the different aspects in the energy deposition in the calorimeter;
   $E_{ECAL}/E_{Total}>0.9$ and $0.7 < E_{Total}/P <1.2$ for electrons and
   $E_{ECAL}/E_{Total}<0.5$ and $E_{Total}/P<0.4$ for muons,
   where $E_{ECAL}$, $E_{Total}$ and $P$ denote the $ECAL$ energy associated with a track,
   total energy deposit in whole calorimeter and track momentum.
   
   \subsubsection{Background Reduction}
   
   After the di-lepton identification,
   we apply the background reduction with following cut variables, which is summarized on the Table~\ref{table:llh}.
   At first we apply the di-lepton mass ($M_{\ell\ell}$) cut which should be consistent with the $Z$ mass:
   $70 < M_{\ell\ell} < 110~{\rm GeV}$ for electron and $70 < M_{\ell\ell} < 100~{\rm GeV}$ for muons, respectively.
   Then we apply the $Z$ flight direction cut: $|\cos{\theta_{Z}}| < 0.8$ to suppress the forward region backgrounds.
   Finally we require the di-jet mass ($M_{jj}$) and recoil mass ($M_{rec}$) cuts to select the Higgs candidate signal:
   $100 < M_{jj} < 140~{\rm GeV},~70 < M_{rec} < 140~{\rm GeV}$ for electron;
   $115 < M_{jj} < 140~{\rm GeV}$ and $70 < M_{rec} < 140~{\rm GeV}$ for muon, respectively.
  \begin{table}[htbp]
   \begin{center}
    \caption{Summary of the background reduction in $\ell^{+}\ell^{-}H$ channel at the $\sqrt{s}=250$ and $350~{\rm GeV}$.}
    \label{table:llh}
    \begin{tabular}{l|r|r|r|r}
     \hline
     $\sqrt{s}$ (GeV)&\multicolumn{2}{c|}{250}&\multicolumn{2}{c}{350}\\
     \hline
     Cut  & Sig. & Bkg. & Sig. & Bkg.\\
     \hline\hline   
     Gen. ($e$)   & 3137 & 4512520 & 2740 & 3822410\\
     Gen. ($\mu$) & 2917 & 4512520 & 1789 & 3822410\\
     \hline
     All cuts ($e$)   & 1184 & 1607 & 567 & 590 \\
     All cuts ($\mu$) & 1365 & 983  & 638 & 465 \\   
     \hline
     Significance (Eff.) ($e$)  & \multicolumn{2}{c|}{22.4 (37.8\%)}& \multicolumn{2}{c}{16.7 (20.7\%)}\\
     Significance (Eff.) ($\mu$)& \multicolumn{2}{c|}{28.2 (46.8\%)}& \multicolumn{2}{c}{19.2 (35.7\%)}\\   
     \hline
    \end{tabular}
   \end{center}
  \end{table}
  
 \section{Measurement Accuracy of Branching Fraction}
 
  \subsection{Template Fitting}
  
  In order to evaluate the measurement accuracies of the Higgs branching fraction (BR), we employ the template fitting method~\cite{template}.
  At first, we prepare the flavor-likeness template samples which is calculated from the LCFIVTX output $x_{1, 2}$:
  \begin{equation}
   x-likeness \equiv \frac{x_{1}x_{2}}{(1-x_{1})(1-x_{2})},
  \end{equation}
  where $x_{1,2}$ represents the flavor tagging output from LCFIVTX for di-jet.
  We assume the Poisson statistics ($P_{ink}$) for each bin $(i,~j,~k)$ of the template samples:
  \begin{equation}
   P_{ijk} = \frac{X^{\mu}e^{-\mu}}{X!}~~~\left(\mu \equiv\sum{N^{template}_{ijk},~X\equiv N^{data}_{ijk}}\right),
  \end{equation}
  where $N^{data}_{ijk}$ is the number of entries in $(i,~j,~k)$ bin.
  $N^{template}_{ijk}$ represents the sum of the number of entries at bin $(i,~j,~k)$ in each template sample:
  \begin{equation}
   N^{template}_{ijk} = \sum_{s=bb,cc,gg,bkg}r_{s}\cdot N^{s}_{ijk},
  \end{equation}
  where $N^{s}_{ijk}$ represents the number of entries at the $(i,~j,~k)$ bin in $H\to b\bar{b},~c\bar{c},~gg$ and background template sample ($N^{bkg}_{ijk}$), which includes the number of entries of SM background and Higgs to none hadronic decays.
  $r_{s}$ represents the fitting parameters of $r_{bb}$, $r_{cc}$, $r_{gg}$ and $r_{bkg}$,
  where they are the ratios of number of entries in Higgs hadronic decays of $H \to b\bar{b},~c\bar{c}$ and $gg$
  after the background reduction to the entries predicted from the SM Higgs BR.
  $r_{bkg}$ is a normalization factor for the SM background and other Higgs none hadronic decays,
  which is fixed to be 1 from the assumption that the SM backgrounds are well understood.
  Finally we apply the template fitting with minimizing the following log-likelihood variable $L$
  calculated from the product of the probability $P_{ijk}$ in each bin:
  \begin{equation}
   L=-\log{\left(\prod_{i,j,k}P_{ijk}\right)}=-\sum_{i,j,k}\left(\log{P_{ijk}}\right).
  \end{equation}
  
  \subsection{Measurement accuracies of BR}
  
  To evaluate the measurement accuracies of the sigma times BRs for $bb$, $cc$ and $gg$,
  we apply the 1000 times template fitting Toy-MC and obtain the fitted results of $r_{s}$ ($s=bb,~cc,~gg$):
  \begin{equation}
   \sigma_{ZH}\cdot BR(H\to s) = r_{s} \times {\sigma_{ZH}}^{SM} \cdot BR(H\to s)^{SM},\label{eq:sigmaxBR}
  \end{equation}
  where $\sigma_{ZH}$ is a Higgs production cross section,
  ${\sigma_{ZH}}^{SM}$ and $BR(H\to s)^{SM}$ are cross section and BR of $H \to s$ predicted in SM.
  From the Eq.~\ref{eq:sigmaxBR}, we obtain the measurement accuracies of Higgs BR from following equation:
  \[
  \displaystyle\frac{\Delta \sigma BR(H\to s)}{\sigma BR} = \sqrt{\left(\frac{\Delta r_{s}}{r_{s}}\right)^{2}+\left(\frac{\Delta \sigma_{ZH}}{\sigma_{ZH}}\right)^{2}}, 
  \]
  here we assume the 2.5\% of cross section measurement uncertainty ($\Delta \sigma_{ZH}/\sigma_{ZH}$)
  estimated from the recoil mass study~\cite{ZHrecoil}.
  Summary tables of the measurement accuracies of Higgs BR are shown in Table~\ref{table:template_250} and \ref{table:template_350} for $\sqrt{s}=250$ and 350 GeV, respectively.
  
  \begin{table}[htbp]
   \begin{center}
    \caption{Summary of template fitting results and BR measurement accuracies at the $\sqrt{s}=250~{\rm GeV}$.}
    \label{table:template_250}
    \begin{tabular}{c|c|c|c|c|c}
     \hline
     & $\nu\bar{\nu}H$ & $q\bar{q}H$ & $e^{+}e^{-}H$ &$\mu^{+}\mu^{-}H$ & comb.\\
     \hline\hline
     $r_{bb}$ & 1.00$\pm$0.016 & 1.00$\pm$0.015 & 1.00 $\pm$ 0.039 & 1.00 $\pm$ 0.33 & 1.00$\pm$0.012\\
     \hline
     $r_{cc}$ & 1.00$\pm$0.12  & 1.00 $\pm$0.12 &  0.98 $\pm$ 0.29 & 1.01 $\pm$ 0.24 & 1.00$\pm$0.09 \\
     \hline
     $r_{gg}$ & 0.99$\pm$0.14  & 1.00$\pm$0.13  &  0.99 $\pm$ 0.35 & 1.00 $\pm$ 0.21 & 1.00$\pm$0.10 \\
     \hline
     $\sigma BR(bb)/\sigma^{SM}$ (\%) & 65.7$\pm$1.1 & 65.7$\pm$1.0  & 65.7 $\pm$ 2.6  & 65.7 $\pm$ 2.2  & 65.7$\pm$0.7 \\
     \hline
     $\sigma BR(cc)/\sigma^{SM}$ (\%)&3.59$\pm$0.43  & 3.61$\pm$0.44 & 3.53 $\pm$ 1.03 & 3.63 $\pm$ 0.85 & 3.60$\pm$0.31\\
     \hline
     $\sigma BR(gg)/\sigma^{SM}$ (\%)&5.46$\pm$0.76  & 5.48$\pm$0.76 & 5.45 $\pm$ 1.94 & 5.49 $\pm$ 1.14 & 5.47$\pm$0.54\\
     \hline\hline
     $\Delta \sigma BR(bb)/\sigma BR$ (\%) & 3.0  &  2.9 &  4.7 &  3.3 & 2.7 \\
     \hline				      
     $\Delta \sigma BR(cc)/\sigma BR$ (\%) & 12.2 & 12.3 & 29.3 & 23.5 & 8.9 \\
     \hline				      
     $\Delta \sigma BR(gg)/\sigma BR$ (\%) & 14.2 & 14.1 & 35.6 & 20.7 & 10.2\\
     \hline
    \end{tabular}
   \end{center}
  \end{table}

 \begin{table}[htbp]
  \begin{center}
   \caption{Summary of template fitting results and BR measurement accuracies at the $\sqrt{s}=350~{\rm GeV}$.}
   \label{table:template_350}
    \begin{tabular}{c|c|c|c|c|c}
     \hline
     & $\nu\bar{\nu}H$ & $q\bar{q}H$ & $e^{+}e^{-}H$ &$\mu^{+}\mu^{-}H$ & comb.\\
     \hline\hline
     $r_{bb}$ & 1.00$\pm$0.012 & 1.00$\pm$0.015 & 1.00 $\pm$ 0.056 & 1.00 $\pm$ 0.051 & 1.00$\pm$0.010 \\
     \hline					       
     $r_{cc}$ & 1.00$\pm$0.10  & 0.99$\pm$0.11  & 1.02 $\pm$ 0.26  & 1.02 $\pm$ 0.32  & 1.00$\pm$0.07 \\
     \hline					       
     $r_{gg}$ & 1.00$\pm$0.10  & 1.00$\pm$0.13  & 0.97 $\pm$ 0.35  & 0.97 $\pm$ 0.35  & 1.00$\pm$0.08 \\
     \hline
     $\sigma BR(bb)/\sigma^{SM}$ (\%) & 65.7$\pm$0.8  & 65.7$\pm$1.0   & 65.6 $\pm$ 3.68 & 65.6 $\pm$ 3.32 &65.7$\pm$ 0.6\\
     \hline
     $\sigma BR(cc)/\sigma^{SM}$ (\%) & 3.60$\pm$0.35 & 3.68$\pm$0.26  & 3.68 $\pm$ 0.94 & 3.66 $\pm$ 1.16 &3.59$\pm$0.26\\
     \hline
     $\sigma BR(gg)/\sigma^{SM}$ (\%) & 5.48$\pm$0.53 & 5.49$\pm$ 0.74 & 5.32 $\pm$ 1.91 & 5.35 $\pm$ 1.94 & 5.48$\pm$ 0.43\\
    \hline\hline
     $\Delta \sigma BR(bb)/\sigma BR$ (\%)  & 2.8  & 2.9  &  6.1 &  5.6 & 2.7\\
     \hline					   
     $\Delta \sigma BR(cc)/\sigma BR$ (\%)  & 10.1 & 11.2 & 25.6 & 31.7 & 7.7\\
     \hline					   
     $\Delta \sigma BR(gg)/\sigma BR$ (\%)  & 9.9  & 13.7 & 36.0 & 36.3 & 8.2 \\
     \hline
    \end{tabular}
  \end{center}
 \end{table}
 
 \section{Extrapolate to the different Higgs masses}
 
\begin{figure}[htbp]
  \begin{minipage}{0.5\textwidth}
   \begin{center}
    \includegraphics[width=\textwidth]{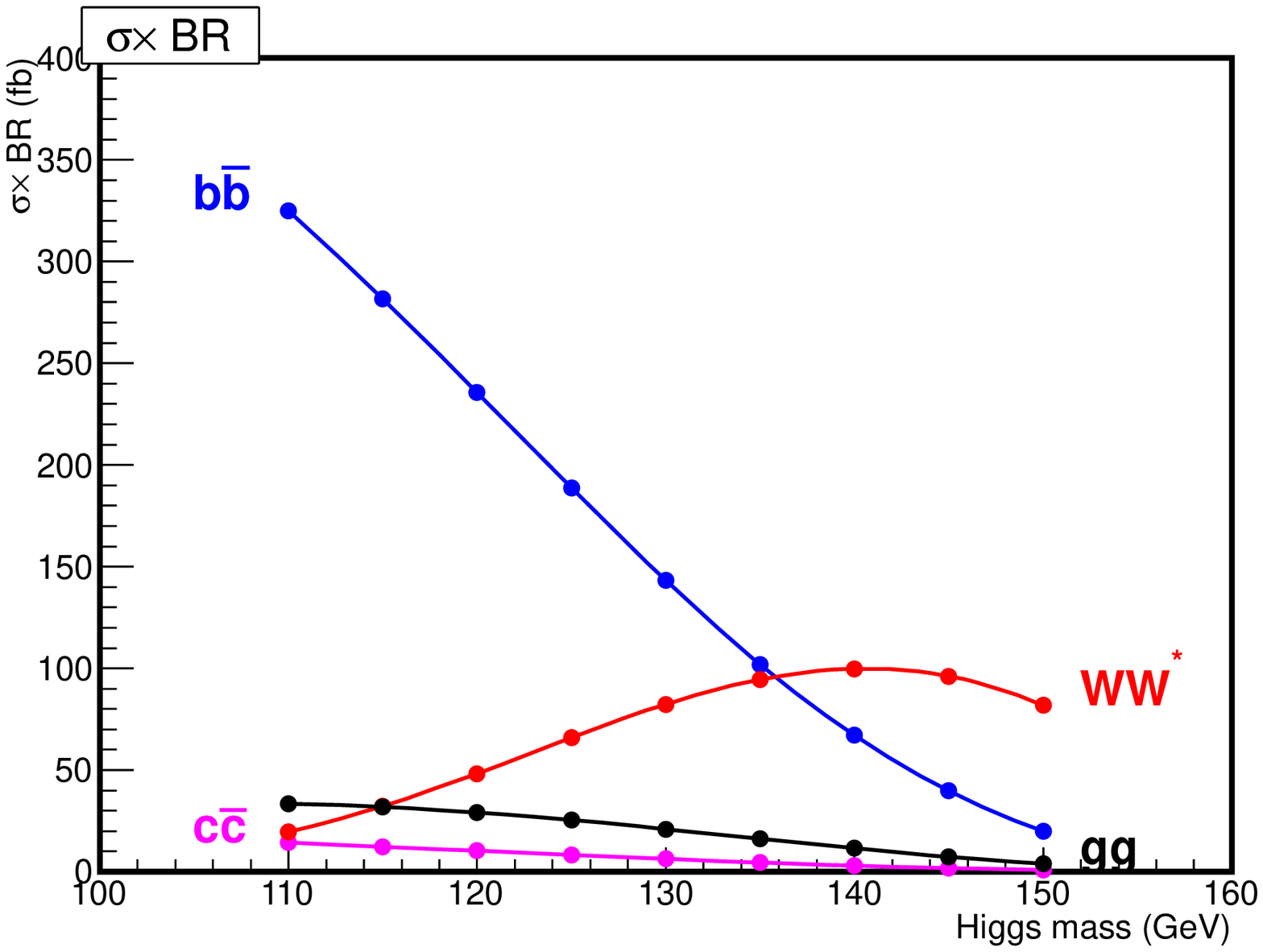}\\   
    (a)
   \end{center}
  \end{minipage}
  \begin{minipage}{0.5\textwidth}
   \begin{center}
    \includegraphics[width=\textwidth]{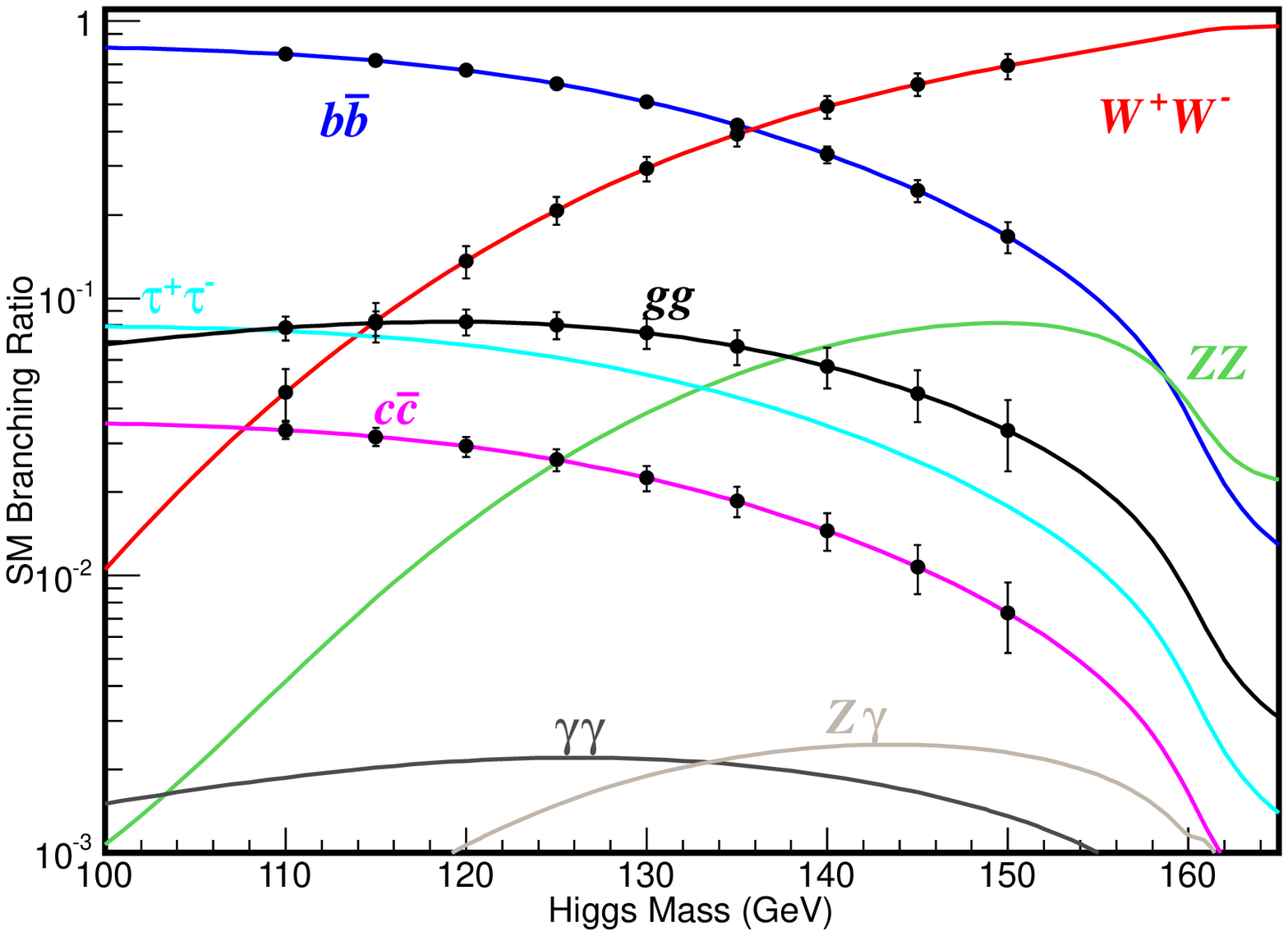}\\    
    (b)
   \end{center}
  \end{minipage}
  \caption{Higgs mass dependence of (a) cross section times Higgs BR and (b) measurement accuracies of the Higgs BR extrapolated from the result at $M_{H}=120~{\rm GeV}$.}
  \label{fig:BRextra}
 \end{figure}

 Since latest LHC results indicate the light Higgs ($115 \leq M_{H} \leq 140~{\rm GeV}$),
 we should extend the results at the $M_{H}=120~{\rm GeV}$ to the prospective mass region.
 In order to evaluate the Higgs BR measurement accuracy at other Higgs masses,
 we extrapolate the results at the $M_{H}=120~{\rm GeV}$ with Eq.~\ref{eq:BRextra}
 while we assume the same background reduction efficiency for other mass. 
 \begin{equation}
  \left( \frac{\Delta \sigma BR}{\sigma BR}(s) \right)_{M_{H}} = \left( \frac{\Delta \sigma BR}{\sigma BR}(s) \right)_{120} \times \sqrt{ \frac{ \sigma_{120}\cdot BR(s)_{120}}{\sigma_{M_{H}}\cdot BR(s)_{M_{H}} } },\label{eq:BRextra}
 \end{equation}
 where $\sigma_{M_{H}}$ and $BR(s)_{M_{H}}$ denote the cross section and BR of $H \to s$ at $M_{H}$,
 as shown in Fig.~\ref{fig:BRextra} (a).
 In addition, we also compile the $H\to WW^{*}\to 4j$ result~\cite{H2WW},
 even though this study assume the electron right-handed polarization: $(e^{+}, e^{-})=(-30\%, +80\%)$
 to suppress the $e^{+}e^{-}\to W^{+}W^{-}$ backgrounds. 
 Figure~\ref{fig:BRextra} (b) shows the expected measurement accuracies of the Higgs BR at each mass
 and summarized on the Table~\ref{table:BRextra}.
  
 \begin{table}[htbp]
  \begin{center}
   {\small
   \caption{Summary of the extrapolated accuracies of the Higgs BR from the result at the $M_{H}=120~{\rm GeV},~\sqrt{s}=250~{\rm GeV}$ to prospective masses calculated by HPROD~\cite{HPROD} and HDECAY~\cite{HDECAY}. Here 2.5\% of the $\sigma_{ZH}$ uncertainty is also included.}
   \label{table:BRextra}
   \begin{tabular}{l|r|r|r|r|r|r|r|r|r}
    \hline
    $M_{H}$ (GeV) & \multicolumn{3}{c|}{120} & \multicolumn{3}{c|}{130} & \multicolumn{3}{c}{140}\\
    \hline
    $\sigma$ ($\rm fb^{-1}$) & \multicolumn{3}{c|}{354.3} & \multicolumn{3}{c|}{279.9} & \multicolumn{3}{c}{203.1}\\
    \hline\hline
     Modes & $BR(\%)$  & $\sigma BR$ & $\displaystyle\frac{\Delta \sigma BR}{\sigma BR}$ & $BR(\%)$ & $\sigma BR$& $\displaystyle\frac{\Delta \sigma BR}{\sigma BR}$ & $BR(\%)$ & $\sigma BR$ & $\displaystyle\frac{\Delta \sigma BR}{\sigma BR}$\\
    \hline
    $H\to b\bar{b}$ & 66.5 & 235.6 & 2.7\% & 51.2 & 143.3 & 3.5\%& 33.0 & 67.1 & 5.1\% \\
    \hline
    $H\to c\bar{c}$ & 2.9 & 10.4 & 8.1\% & 2.3 & 6.3 & 10.4\%& 1.5 & 3.0 & 15.2\%\\
    \hline
    $H\to gg$ & 8.2 & 29.2 & 9.0\%  & 7.5 & 21.0 & 10.6\%& 5.7 & 11.5 & 14.3\%\\
    \hline
    $H\to W^{+}W^{-}$ & 13.6 & 48.3 & 15.7\% & 29.4 & 82.4 & 10.3\%& 49.2 & 99.8 & 9.3\%\\
    \hline
   \end{tabular}
   }
  \end{center}
 \end{table}
 
 \section{Conclusion}

 We evaluate the measurement accuracies of the Higgs BR for $H \to b\bar{b}$, $c\bar{c}$ and $gg$ channels.
 With the template fitting analysis, we obtain the measurement accuracies of $\sigma \times BR$ for $H \to b\bar{b}$, $c\bar{c}$ and $gg$ as 3\%, 9\% and 10\%, respectively.
 We also estimate the Higgs BR measurement accuracies at the prospective mass region by LHC
 with extrapolating the results at the Higgs mass of 120 GeV.

 \section*{Acknowledgments}
 
 We would like to acknowledge the members who join the ILC physics WG subgroup~\cite{physwg}
 for useful discussion of this work and to ILD analysis group members who maintain the software and MC samples for this work.
 This work is supported in part by Creative Scientific Research Grant No. 18GS0202 of the Japan Society for Promotion of
 Science (JSPS), the JSPS Core University Program, and the JSPS Grant-in-Aid for Scientific Research No. 22244031.

 \section{Bibliography}
 
 \begin{footnotesize}
  
 \end{footnotesize}
 

\end{document}